\documentclass[aps,prb,showpacs,superscriptaddress,amsmath,twocolumn]{revtex4}

\usepackage{amsmath}
\usepackage{amssymb}
\usepackage{graphicx}

\DeclareMathOperator{\Exp}{Exp}

\begin{document}

\title{Exploring the graphene edges with coherent electron focusing}

\author{P. Rakyta}
\affiliation{Department of Physics of Complex Systems,
E{\"o}tv{\"o}s University,
H-1117 Budapest, P\'azm\'any P{\'e}ter s{\'e}t\'any 1/A, Hungary
}

\author{A.~Korm\'anyos}
\thanks{e-mail: a.kormanjos@lancaster.ac.uk}
\affiliation{Department of Physics, Lancaster University,
Lancaster, LA1 4YB, UK}

\author{J. Cserti}
%\thanks{e-mail: cserti@complex.elte.hu}
\affiliation{Department of Physics of Complex Systems,
E{\"o}tv{\"o}s University,
H-1117 Budapest, P\'azm\'any P{\'e}ter s{\'e}t\'any 1/A, Hungary
}

\author{P. Koskinen}
\affiliation{Department of Physics, NanoScience Center, 40014
University of Jyv\"askyl\"a, Finland
}

%\wideabs{

\begin{abstract}
 We study theoretically the coherent electron focusing in graphene nanoribbons.
Using semiclassical and numerical tight binding calculations  we show that  armchair
edges give rise to equidistant peaks in the focusing spectrum. In the case of zigzag
edges at low magnetic fields one can also observe focusing peaks but  
 with increasing magnetic field a more complex interference structure emerges in
the spectrum.
 This difference in the spectra can be observed even if the zigzag edge
undergoes structural
 reconstruction. Therefore transverse electron focusing can help in the identification and
characterization of the edge structure of graphene samples. 
\end{abstract}


\maketitle

\section{Introduction}
Transverse electron focusing (TEF) is a versatile experimental technique which
has been used in metals to study 
the shape of the Fermi surface and the scattering on various surfaces and
interfaces\cite{tsoi}. 
The accessibility of  the quantum ballistic transport regime in two dimensional
electron gas (2DEG) 
opened up the way to the experimental demonstration of the 
\emph{coherent electron focusing}\cite{Houten_Carlo1:cikk}  in GaAs
heterostructures as well.

\begin{figure}[thb]
\includegraphics[scale=0.35]{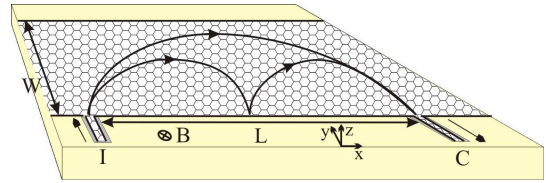}
\caption{(Color online) Schematic geometry  of the transverse electron focusing
setup. A graphene nanoribbon is contacted by an injector (I) and a collector (C) 
probe and  perpendicular (to the graphene sheet) magnetic field  is applied. 
Classical quasiparticle trajectories leaving from the injector at normal
direction, depending on the magnetic field, can  be focused onto the collector.
\label{fig:geometry-focusing}} 
%\vspace{-5mm}
\end{figure}

The  geometry of the coherent electron focusing %studied  in
Ref.~\onlinecite{Houten_Carlo1:cikk} 
is shown in Fig.\ref{fig:geometry-focusing}. The current is injected into the 
sample at  a quantum point contact called injector (I) in perpendicular magnetic
field.  
If the magnetic field is an integer multiple of a focusing field $B_{focus}$, 
electrons injected within a small angle around the perpendicular direction to
the sample edge can be focused onto a second quantum
point contact (the collector, denoted by $C$ in
Fig.~\ref{fig:geometry-focusing}) 
which acts as a voltage probe. Therefore, if  the collector voltage is plotted
as a function of  magnetic field one can observe equidistant peaks at magnetic fields
$B=p*B_{focus}$ ($p=1,2,3,\dots$). The first focusing peak corresponds to
electrons reaching the collector directly, i.e. without bouncing off the edge (see
Fig.~\ref{fig:geometry-focusing}). 
Subsequent peaks correspond to trajectories
bouncing off the edge $p-1$ times before reaching the collector 
and therefore their presence can attest to the specular nature of 
the scattering at the edge.  In other words, the $p \ge 2$ focusing peaks  can
give information on the scattering process taking place at the edge of the sample.

The edge structure of graphene\cite{graphene} 
nanoribbons\cite{han,dresselhaus,zettl,tapaszto} and  graphene
flakes\cite{ritter} 
have recently attracted a lot of interest because it strongly influences
the nanoribbons' and flakes' electronic and  magnetic
 properties\cite{wakabayashi,wakabayashi-2,louie,pisani,koskinen,wassmann}. 
Theoretically, the most often studied edge structures are the armchair and
zigzag ones which have recently  been observed experimentally as
well\cite{dresselhaus,zettl,tapaszto}. 
Density functional calculations suggested  that other types of edges might also
be present, comprising pentagons and heptagons of carbon atoms\cite{koskinen}. 
Experimental evidence for this type of 
edge reconstruction has indeed been found very recently\cite{koskinen-2}.  
The effect of the hydrogen concentration of the 
environment on the edges has also been studied and further possible edge
structures identified\cite{wassmann}. 
Experimentally however the identification and characterisation of the edge
structure has often been a challenge\cite{dresselhaus}. 

In this paper we show that coherent electron focusing 
can  be used in ballistic graphene samples to study the properties of  the edge
structure.  We argue that in the case of armchair edges one would see equidistant peaks in
the focusing spectrum at integer multiples of a focusing field $B_{focus}$. 
In contrast, for zigzag and reconstructed zigzag\cite{koskinen} (reczag) edges only  the first few focusing 
peak would be identifiable 
and for stronger magnetic fields a more complex interference structure would appear in the
focusing spectrum. The presence or absence of focusing peaks at stronger magnetic fields can
therefore discriminate between armchair and zigzag (reczag) edges.

\begin{figure}
\includegraphics[width=85mm]{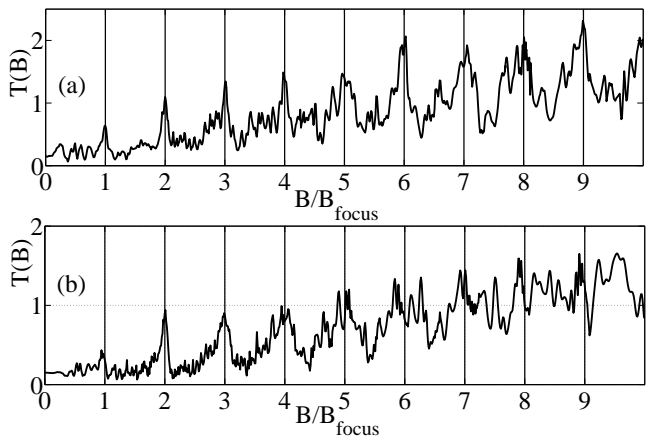}
\caption{The results of tight binding calculations for the transmission probability $T(B)$ 
 as a function of  magnetic field. (a) for an armchair; (b) for a zigzag nanoribbon. 
The black vertical lines indicate the positions of the focusing peaks predicted
by the semiclassical theory, see Sections \ref{sec:semiclass} and \ref{sec:focusing}.
\label{fig:focusing-main}} 
%\vspace{-5mm}
\end{figure}

Our main results are summarized in Fig.~\ref{fig:focusing-main}.
Using the tight-binding model for graphene  with 
nearest-neighbour hopping we numerically calculated the transmission probability
$T(B)$ from the injector to the collector as a function of the magnetic field $B$ 
for three different types of graphene nanoribbons: armchair, zigzag and zigzag
with reconstructed edges, denoted by zz(57) in Ref.~\onlinecite{koskinen}.
As Fig.~\ref{fig:focusing-main} shows  one can indeed observe peaks in the focusing 
spectrum of these nanoribbons. (In the one orbital per site approximation we used
in the computations the focusing spectrum of  the reczag nanoribbon  is 
very similar to %almost indistinguishable from 
the simple zigzag one's therefore we  only show  the latter here.)
In the case of the reconstructed edge  we  took into account that the hopping
between the atoms is different on the heptagons and pentagons than in the bulk 
of the graphene. To obtain realistic nearest-neighbour hoppings we employed 
\emph{ab-initio} calculations. 
(See Section \ref{sec:num-calc} for the details of the \emph{ab-initio} method 
and Ref.~\onlinecite{parameters} for the actual parameters of the calculations). 
We consider the case where there is a finite  carrier density in the sample,
 i.e. the Fermi energy is well above of the Dirac point (this ensures that the semiclassical 
approach we take in Sections \ref{sec:semiclass} and \ref{sec:focusing} is justified). 
We will  focus on the  transmission peaks which can be observed for 
$B/B_{focus}\gtrsim 1$ because they can give local information on one of the edges 
(see Section \ref{sec:semiclass}). 
They should  also be present if e.g. a graphene
flake is contacted  by two probes, as long as the edge between the probes is not
disordered. The first peak in the transmission for 
 all three types of nanoribbons can be found at $B/B_{focus}\approx 1$ where
$B_{focus}=\frac{2\hbar k_F}{e L}$ 
($k_F$ is the Fermi wavenumber (measured from the $\boldsymbol{K}$ point)
and $L$ is the distance between the injector and the collector).
While for armchair edge there are well-defined peaks whenever $B$ is integer
multiple of $B_{focus}$ [see Fig.~\ref{fig:focusing-main}(a)],  in the case of  
zigzag edges (both the ideal and the 
reconstructed one)  a more  complex interference pattern emerges for $B/B_{focus}\gtrsim 6$ 
showing many oscillations but not a clear peak structure [Fig.~\ref{fig:focusing-main}(b)].  
These results imply that a) one can distinguish the armchair and zigzag edges by their high magnetic 
field focusing spectrum; b) the difference in the focusing spectra  persists even if the  zigzag edge 
 undergoes structural reconstruction. 
We will explain the differences in the focusing spectra of armchair and zigzag
nanoribbons making use of the semiclassical theory of graphene, introduced in
Refs.~\onlinecite{semi_rival,semi_sajat}.

The rest of the paper is organized as follows. First, in Section
\ref{sec:semiclass} we discuss
the boundary conditions for the semiclassical theory %\cite{semi_rival,semi_sajat}
on the edges of armchair and zigzag graphene nanoribbons and derive the quantization condition for
edge states in magnetic field. The quantization condition then allows us to 
calculate the band structure. 
Using this  in Section \ref{sec:focusing} we show that the focusing spectra
of armchair and zigzag nanoribbons are different in strong magnetic fields.
We then turn to the comparison of the results of numerical calculations and the theoretical
predictions. In Section \ref{sec:num-calc} we discuss some  details of the numerical calculations 
which underpin the theoretical approach presented in Sections \ref{sec:semiclass} and \ref{sec:focusing}. 
Finally, in Section \ref{sec:summary} we give our conclusions.

\section{Semiclassical theory of edge states in graphene}
\label{sec:semiclass}
We start our discussion by establishing the link between the theory of
semiclassical approximations for
graphene  and the boundary conditions for Dirac fermions on honeycomb lattice.
 If the magnetic length $l_B=\sqrt{\hbar/{|eB|}}$ is much larger than the
lattice constant of graphene, 
the general energy-independent boundary condition 
has the form of a local restriction on the components of the wave function
$\Psi$
at the edge ($\mathcal{E}$)\cite{perem_Falko,perem_Carlo}.  It can be cast into
the following form:
$\hat{M}\Psi = \Psi$  
where the $4\times4$ matrix $\hat{M}$  may be chosen as Hermitian and unitary
matrix:
$\hat{M} = \hat{M}^{\dag}$, 
and $\hat{M}^2 = \hat{I}$. 
One can show\cite{perem_Carlo} that demanding: a) that the probability current
normal to the boundary be zero;
b) that the boundary should preserve the electron-hole symmetry of the bulk; c)
and finally, assuming 
 that the boundary conditions do not break the time reversal symmetry, leads to
the following form of 
the matrix  $\hat{M}$:
$ \hat{M} =\big(
\boldsymbol{\nu}\boldsymbol{\tau}\otimes\boldsymbol{n}\boldsymbol{\sigma}\big)$,
where $\boldsymbol{\sigma}=(\sigma_x,\sigma_y,\sigma_z)$,
$\boldsymbol{\tau}=(\tau_x,\tau_y,\tau_z)$ and 
$\sigma_i$, $\tau_i$ are Pauli matrices acting in the 
sublattice and valley space, respectively. Furthermore, 
$\boldsymbol{\nu}$ and $\boldsymbol{n}$ are three dimensional unit vectors, 
restricted to two classes: zigzag-like ($\boldsymbol{\nu}=\pm\hat{\mathbf{z}}$, 
$\mathbf{n}=\hat{\mathbf{z}}$, where $\hat{\mathbf{z}}$ is the unit vector
perpendicular to the plane of the 
graphene sheet) and armchair-like ($\nu_z=n_z=0$). 
Additionally in both classes $\mathbf{n}\perp\mathbf{n}_{\mathcal{E}}$, where 
$\mathbf{n}_{\mathcal{E}}$ is a  unit vector in the plane of the graphene sheet
and it is 
perpendicular to the edge.

Let us  now consider the reflection from an edge of a nanoribbon in more
details. 
It follows from the form of the boundary conditions described earlier 
that the wavefunctions $\Psi^{\pm}_{\mathcal{E}}$ 
on the boundary is proportional to the 
eigenvectors $\boldsymbol{Z}^{\pm}$ corresponding to the doubly degenerate
 unit eigenvalue of the matrix $\hat{M}$. In other words, 
$\Psi^{\pm}_{\mathcal{E}}=\eta^{\pm}\boldsymbol{Z}^{\pm}e^{{\rm i}kx}$ where
$\Psi^{+}$ ($\Psi^{-}$) and 
$\boldsymbol{Z}^{+}$ ($\boldsymbol{Z}^{-}$) correspond to isospin vector
$\boldsymbol{\nu}$ (-$\boldsymbol{\nu}$),
$\eta^{\pm}$ are amplitudes and $k_x$ is the wavevector component along the
(translationally invariant) edge.
On the other hand, one can show that in  magnetic fields where $l_B\gg
\lambda_F=2\pi/k_F$  
($\lambda_F$ is the Fermi wavelength) the wavefunctions
$\Psi^{\pm}_{\mathcal{E}}$  
can also  be written as a superposition of an incident $\Psi_{in}$ 
and reflected $\Psi_{out}$ 
plane wave:
$\Psi^{\pm}_{\mathcal{E}}=\Psi^{\pm}_{in}+\hat{r}^{\pm}\Psi^{\pm}_{out}$
where $\hat{r}^{\pm}$ are reflection amplitudes. By equating the two forms of
$\Psi^{\pm}_{\mathcal{E}}$
the coefficients $\hat{r}^{\pm}$ and $\eta^{\pm}$ can be easily obtained for any
boundary condition described by the 
matrix $\hat{M}$.
For instance, in the case of armchair edge,  where the isospin vector can be 
parametrized as  $\boldsymbol{\nu} = (\cos\varphi,\sin\varphi,0)$ and
$\mathbf{n}=(1,0,0)^{T}$,
the eigenvectors of $\hat{M}$ with unit eigenvalues are 
\begin{equation}	
\boldsymbol{Z}^{\pm}_a = \frac{1}{2}\begin{pmatrix}
                                   e^{-{\rm i}\frac{\varphi}{2}} \begin{pmatrix}
1 \\ \pm 1  \end{pmatrix}\\
                                   e^{{\rm i}\frac{\varphi}{2}} \begin{pmatrix}
\pm1 \\ 1  \end{pmatrix}
                                   \end{pmatrix}. 
\label{eq:Z}
\end{equation}
The ansatz for  $\Psi^{\pm}_{\mathcal{E}}$ can also be written as  
\begin{equation}
 \Psi^{\pm}_{\mathcal{E}} = \left[\frac{1}{2}\begin{pmatrix}
                                   e^{-{\rm i}\frac{\varphi}{2}} \begin{pmatrix}
e^{{\rm i}\alpha} \\ 1  \end{pmatrix}\\
                                    \pm e^{{\rm
i}\frac{\varphi}{2}}\begin{pmatrix} e^{{\rm i}\alpha} \\ 1  \end{pmatrix}
                                   \end{pmatrix} +
\frac{\hat{r}^{\pm}}{2}\begin{pmatrix}
                                   e^{-{\rm i}\frac{\varphi}{2}} \begin{pmatrix}
e^{-{\rm i}\alpha} \\ 1  \end{pmatrix}\\
                                    \pm e^{{\rm
i}\frac{\varphi}{2}}\begin{pmatrix} e^{-{\rm i}\alpha} \\ 1  \end{pmatrix}
                                   \end{pmatrix}\right] e^{{\rm i}kx}\;
\label{eq:felbontottPsi}
\end{equation}
where $\alpha$ is the incidence angle (measured from the $x$ axis, see
Fig.~\ref{fig:geometry-focusing}
). 
Equating the two forms of $\Psi^{\pm}_{\mathcal{E}}$, i.e.
 $\Psi^{\pm}_{\mathcal{E}}=\eta^{\pm}\boldsymbol{Z}^{\pm}e^{{\rm i}kx}$ and
Eq.~(\ref{eq:felbontottPsi})
 a straightforward calculation gives
\begin{equation}
\hat{r}^{\pm}_a = 
\Exp({\rm i}\Delta\Phi^{\pm}_a); \quad\mbox{where}\quad
\Delta\Phi^{+}_a=\alpha,\,\, \Delta\Phi^{-}_a=\alpha+\pi.
\label{eq:ref_coef-armchair}
\end{equation}

In the case of zigzag edges the boundary can be characterized by a superlattice
vector 
$\boldsymbol{T}=m\boldsymbol{a}_1+l\boldsymbol{a}_2$ where $m\neq l$  are
integers and  
$\boldsymbol{a}_1$, $\boldsymbol{a}_2$ are the lattice vectors of
graphene\cite{perem_Carlo}. 
The boundary condition is given by $\hat{M}=\xi_{ml}\left(\tau_z\ \otimes
\sigma_z\right)$
where  $\xi_{ml}=\textnormal{sgn}(m-l)$, where $\textnormal{sgn}(\dots)$ is the
sign function. 
After performing analogous calculations as for the armchair edge one finds that 
\begin{equation}
\hat{r}^{\pm}_z=\Exp({i\Delta\Phi^{\pm}_z});\quad\mbox{where}\quad
\Delta\Phi^{\pm}_z=\alpha-\pi\mp\xi_{ml}\alpha.
\label{eq:ref_coef-zigzag}
\end{equation}

The physical meaning of Eqs.~(\ref{eq:felbontottPsi}),
(\ref{eq:ref_coef-armchair}) and 
(\ref{eq:ref_coef-zigzag}) is the following:  
assuming specular reflection 
in classical picture the momentum of the particle 
is rotated upon reflection. Since the quasiparticles in graphene are chiral, the
change in the direction of 
their momentum rotates the  pseudospin as well and therefore it leads to 
a change in the phase of their wavefunction.
As one can see from Eqs.~(\ref{eq:ref_coef-armchair}) and
(\ref{eq:ref_coef-zigzag}) this phase shift  depends
on the type (armchair v. zigzag) of the edge as well.  
This will be  important when we use semiclassical quantization to obtain the
band structure of 
 graphene armchair and zigzag nanoribbons. 

A second ingredient in the calculation of the edge states is the semiclassical
theory  
introduced in  Refs.~\onlinecite{semi_rival,semi_sajat}. As 
it has been shown\cite{semi_rival,semi_sajat} %in
Refs.~\onlinecite{semi_rival,semi_sajat}, 
one can introduce a classical 
Hamiltonian $\mathcal{H}_e(\mathbf{p},\mathbf{r})=v_F\sqrt{\left(
\mathbf{p}-e\mathbf{A}\right)^2} + V(\mathbf{r})$ 
for graphene to describe the classical motion of electron-like quasiparticles. 
Here $\mathbf{p}=(p_x,p_y)$ is the canonical momentum, 
$\mathbf{A}(\mathbf{r})$ is the vector potential describing any external
magnetic field
 and $V(\mathbf{r})$ is  scalar potential 
which is taken to be zero throughout our discussion, $V(\mathbf{r})=0$. If the
edges of the graphene 
ribbon do not break 
the translational invariance the system is integrable because the longitudinal
momentum $p_x$ and 
the energy are conserved (see Fig.~\ref{fig:geometry-focusing} for the choice of
the coordinate system). 
Using the Landau gauge  $\mathbf{A}(y)=(By,0,0)^T$ for the vector potential to
describe perpendicular magnetic 
field pointing into the $-\hat{z}$ direction, 
the Hamiltonian $\mathcal{H}_e$ will not depend on the $x$ coordinate and the 
only nontrivial quantization condition is related to the motion perpendicular to
the graphene edges 
(in the $\hat{y}$ direction). In general it can be written as 
\begin{equation}
\frac{1}{\hbar}S_y + \gamma + \Delta\Phi_{\mathcal{E}}^{\pm}=
2\pi\left(n^{\pm}+\frac{\mu}{4}\right)\;,
\label{eq:EBK_1D}
\end{equation}
where $S_y = \oint p_y\,dy$ is the classical action and   the integration is
over one period of the motion 
perpendicular to the edges.
Furthermore, $\Delta\Phi_{\mathcal{E}}^{\pm}$ are the phase shifts coming from
the reflections at the edges; 
$n^{\pm}$ is a positive integer, $\mu$ is the Maslov index counting the number
of caustic points  and finally 
$\gamma$ is a  Berry-phase-like quantity \cite{semi_rival,semi_sajat} which  can
be calculated from the equation
 $\frac{{\rm d}}{{\rm d}t}\gamma\left(\mathbf{r}(t)\right) = 
\frac{1}{2}\left( \nabla_{\mathbf{r}}\times\frac{{\rm d}}{{\rm
d}t}\mathbf{r}\right)_z$. 
Here the classical trajectory $\mathbf{r}(t)$ of quasiparticles is given
by\cite{semi_rival,semi_sajat}
$\frac{{\rm d}}{{\rm d}t}\mathbf{r} =
\nabla_{\mathbf{p}}\mathcal{H}_e(\mathbf{p},\mathbf{r})$.
One can show that in  perpendicular  magnetic field 
$\gamma(\mathbf{r})$ is half of the deflection angle of the momentum of the
quasiparticles. 

In Landau gauge the classical action $S_y$ can be calculated in the same way 
as in Ref.\onlinecite{semi_sajat}.  
The detailed presentation of the semiclassical quantization of armchair and
zigzag nanoribbons 
in perpendicular magnetic field, discussing all the possible classical orbits
and the corresponding 
$\gamma$ and $\Delta\Phi_{\mathcal{E}}^{\pm}$ phases is left for a forthcoming 
publication.
For our purposes it will be  sufficient to consider only those classical 
orbits which correspond to skipping motion along the 
boundaries, see e.g. in Fig.\ref{fig:geometry-focusing}.  If the magnetic field
is strong enough such that
the diameter of the cyclotron orbit is smaller than the width of the nanoribbon 
(denoted by $W$ in Fig.\ref{fig:geometry-focusing}), 
i.e. for $2 R_c< W$ where $R_c=\frac{E_F l_B^2}{\hbar v_F}$ is the cyclotron
radius at Fermi energy $E_F$, one can show that 
only these  skipping orbits correspond to current-carrying states.  The phase
shift 
$\Delta\Phi^{\pm}_{\mathcal{E}}$ coming from the reflection 
at the edge  is given by Eq.~(\ref{eq:ref_coef-armchair}) and
Eq.~(\ref{eq:ref_coef-zigzag}) 
for armchair and zigzag edges, respectively.
Considering first armchair nanoribbons, 
using  Eqs.~(\ref{eq:ref_coef-armchair}) and ~(\ref{eq:EBK_1D})
the semiclassical quantization of skipping orbits reads
\begin{equation}
k_F R_c \left(\frac{\pi}{2} + \beta_n^{\pm}+\frac{1}{2}\sin
2\beta_n^{\pm}\right) =2\pi 
\left(n^{\pm} \pm \frac{1}{4}\right)
\label{eq:edge-quant-armchair}
\end{equation}
where $k_F=\frac{E_F}{\hbar v_F} $, $\beta_n^{\pm}$ is the angle with the $y$
axis under which the cyclotron orbit 
is reflected from the boundary, and $+$($-$) corresponds to isospin
$\boldsymbol{\nu}$($-\boldsymbol{\nu}$).
Moreover, $n^+=0 \dots n_{max}^+$ ($n^-=1 \dots n_{max}^-$) where 
$n_{max}^{\pm}$ is the largest integer smaller than $\frac{1}{2}k_F R_c\mp
\frac{1}{4}$. 
This quantization conditions holds for both metallic and insulating armchair
nanoribbons. 
The physical meaning of Eq.~(\ref{eq:edge-quant-armchair}) is that the enclosed
flux by the cyclotron orbit 
and the edge of the nanoribbon is quantized and  it must equal $h/e(n^{\pm}\pm
1/4)$.  It is interesting to note  that for 2DEG with hard wall boundary conditions 
the quantization condition for edge states looks 
very similar, except that one would have to omit the $\pm$  sign  and on  the
right hand side one would have $(n+\frac{3}{4})$.

The semiclassical quantization for edge states in the case of zigzag nanoribbon
can be obtained in the same way
 as for an armchair nanoribbon, the only difference is that  instead of
Eq.~(\ref{eq:ref_coef-armchair}) one has to use   
Eq.~(\ref{eq:ref_coef-zigzag}). One finds that it is given by
\begin{equation}
\begin{split}
k_F R_c\left(\frac {\pi}{2} +\beta_n^{\pm}+\frac{1}{2}\sin
2\beta_n^{\pm}\right)- &
\nu \xi_{ml} \left( \frac{\pi}{2}+\beta_n^{\pm} \right) \\ = & \,\, 2\pi
\left(n^{\pm}-\frac{1}{4}\right).
\label{eq:edge-quant-zigzag}
\end{split}
\end{equation}
Here $\pm$  corresponds to the product $\nu \xi_{ml}=\pm 1$, where both $\nu$
and $\xi_{ml}$ can take on
values $\pm 1$. The value of $\nu$ depends on the 
 isospin: it is $1$ ($-1$) if the isospin is $\boldsymbol{\nu}=\hat{z}$
($\boldsymbol{\nu}=-\hat{z}$) 
and $\xi_{ml}=\pm 1 $ is introduced before Eq.~(\ref{eq:ref_coef-zigzag}).
The range of the quantum number $n^{\pm}$ is the same as for the armchair case.
One can notice that 
as compared to Eq.~(\ref{eq:edge-quant-armchair}) there is an extra term on the
left hand side of 
Eq.~(\ref{eq:edge-quant-zigzag}). The origin of the difference in the focusing
spectra of armchair and zigzag nanoribbons,  to be discussed in  Section
\ref{sec:focusing},  can be traced back to this term in the dispersion relation.

We have calculated the band structure for armchair and zigzag nanoribbons
in homogeneous perpendicular magnetic field both semiclassically, using
Eqs.~(\ref{eq:edge-quant-armchair}),~(\ref{eq:edge-quant-zigzag}),
and numerically in tight binding  approximation.
\begin{figure}[htb]
\includegraphics[width=85mm,height=55mm]{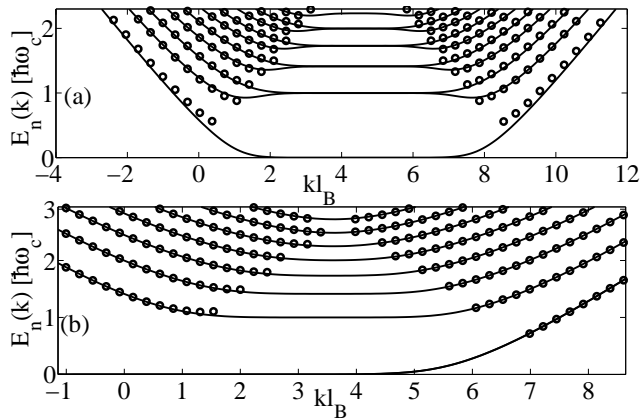}
\caption{Comparison of the results for the band structure obtained from
semiclassical quantization given by Eqs.~(\ref{eq:edge-quant-armchair}),
~(\ref{eq:edge-quant-zigzag}) (circles) and from numerical tight binding
calculations (solid lines). 
The energy is in units of $\hbar\omega_c=\sqrt{2}\frac{\hbar v_F}{l_B}$.
%the wave vector $k$ is in  units of $l_B$. 
The magnetic field is given by $W/l_B=8.97$.
(a) armchair nanoribbon; (b) zigzag nanoribbon, in the vicinity of the
$\boldsymbol{K}$ point.
\label{fig:disp}}
\end{figure}
As one can see  in Fig.~\ref{fig:disp} the agreement between the semiclassical
and the tight-binding calculations is very good except for  low energies $E/
\hbar\omega_c\lesssim 1/2$. 
(The dispersionless state of a zigzag nanoribbon at zero energy cannot be
described by semiclassics either.) 
However, the dispersionless sections of the band structure 
corresponding to Landau levels at finite energies can be calculated
semiclassically\cite{semi_rival,semi_sajat}.
(We note that similar results have been obtained in Ref.~\onlinecite{gusynin}
using the Dirac-like Hamiltonian for graphene.)

\section{Magnetic focusing in graphene nanoribbons}
\label{sec:focusing}

Having obtained the quantization condition for edge states in
Eqs.~(\ref{eq:edge-quant-armchair}) and
~(\ref{eq:edge-quant-zigzag}), the calculation of the magnetic fields
$B_{focus}$, where the transmission 
between the injector and collector is sharply peaked,  follows the reasoning  of
Ref.~\onlinecite{Houten_Carlo1:cikk}. 
The ballistic transport along the edges of a nanoribbon can be understood in
terms of the  
edge states described in Section \ref{sec:semiclass} because they are the
propagating modes of this problem.
If the injector is narrow ($\sim \lambda_F$) one can assume that it excites
these modes coherently. 
Therefore, as long as the distance between the injector and the collector is
smaller than the phase coherence
length, the interference of the edge states can be important.     
For a given $E_F$ the interference of the edge states, labelled by  $n^{\pm}$
in 
Eqs.~(\ref{eq:edge-quant-armchair}) and (\ref{eq:edge-quant-zigzag}), is
determined by the 
phase factors $\exp(i k_n^{\pm} L)$. 
Here the wave numbers $k_n^{\pm}$ are given by $k_n^{\pm}= k_F \sin\beta_n^{\pm}
$  
and $L$ is the distance between the (very narrow) injector and collector. 
Both in the  armchair and in the zigzag case one can show that
 around $n^{\pm}=n_{max}^{\pm}/2$, which corresponds to $|\beta_n^{\pm}|\ll 1$, 
in good approximation the angles $\beta_n^{\pm}$ depend linearly on $n^{\pm}$. 
 
Expanding Eq.~(\ref{eq:edge-quant-armchair}) around $\beta_n^{\pm}=0$ one finds
that 
\begin{equation}
k_n^{\pm} L \approx \frac{\pi L}{R_c}\left(n^{\pm}\pm\frac{1}{4}\right)-C_a+ 
\frac{k_F L}{3}
\left(\frac{\pi}{4}\frac{n_{max}^{\pm}-2n^{\pm}}{n_{max}^{\pm}}\right)^{3}
\label{eq:armchair-foc-cond}
\end{equation}
where $C_a=\frac{\pi}{4}k_F L $. This result means that if $L/2R_c$ is an
integer (or equivalently, 
$B/B_{focus}$ is integer, where $B_{focus}=\frac{2 \hbar k_F}{e L}$)
some  edge channels, with quantum numbers $n^{\pm}$ centred around
$n_{max}^{\pm}/2$,  
can constructively interfere at the collector. 
Other edge states, to which the linear expansion shown in
Eq.~(\ref{eq:armchair-foc-cond}) 
cannot be applied, give rise to an additional interference pattern which does
not have a simple periodicity.
The expression for the focusing field  $B_{focus}$ 
is formally the same as for 2DEG\cite{Houten_Carlo1:cikk} but the dependence of 
$k_F$ on  the electron density  is different in the two systems.

Repeating the expansion for zigzag nanoribbons using
Eq.~(\ref{eq:edge-quant-zigzag}) we find that to linear order 
\begin{equation}
k_n^{\pm} L \approx \frac{1}{1\mp \frac{1}{4\kappa^2}}\frac{\pi
L}{R_c}\left(n^{\pm}+\frac{1}{4}\pm\frac{1}{4}\right)-C_z\\
\label{eq:zizgag-foc-cond}
\end{equation}
where $\kappa=E_F/\hbar\omega_c$ is the filling factor and
$C_z=-\frac{\pi}{4}k_F L \frac{1}{1\mp \frac{1}{4\kappa^2}}$.
For nanoribbons where the cyclotron radius  becomes smaller than the width of
the ribbon
such that $\kappa\gg 1$ is satisfied and therefore  
$1/\left(1\mp \frac{1}{4\kappa^2}\right)\approx 1$  the focusing field
$B_{focus}$ can be defined
in the same way as for armchair ribbons 
and  the transmission peaks are at the same focusing fields  as in the armchair
case. 
%This result is also confirmed by the numerical calculations (see 
%Fig.~\ref{fig:focusing-main}).
However, according to  Eq.~(\ref{eq:zizgag-foc-cond}) 
as $\kappa$ gets smaller for increasing magnetic field (assuming fixed Fermi
energy) the focusing field for the two isospin index becomes slightly different,
moreover, the subsequent focusing fields for each isospin index are no longer equidistant. 
(Examination of the higher order terms not shown in Eq.~(\ref{eq:zizgag-foc-cond}) 
reveals that also the number of interfering edge channels becomes different for 
the two isospin indices.) As a result, a more complex interference pattern is
expected to appear than in the case of armchair nanoribbon and a simple $B_{focus}$  
can no longer be defined.
 
Turning to the comparison of the numerical results shown in Fig.~\ref{fig:focusing-main} 
with the analytical predictions given in 
Eqs.~(\ref{eq:armchair-foc-cond}) and (\ref{eq:zizgag-foc-cond}), one should focus on the 
$B/B_{focus}\ge 1$ regime, where  the edge states are the current carrying modes. 
One can see that the peak positions for the armchair nanoribbon are in very good agreement 
with the prediction of the semiclassical theory. 
Up to $B/B_{focus}\lesssim 5$ the focusing peaks
are  also clearly discernible in the focusing spectra of zigzag nanoribbons  but 
for stronger magnetic fields a more complex interference   pattern emerges. The peak 
at  $B/B_{focus}=7$ for example can clearly be seen for the armchair case. In
contrast, for the zigzag  
edge  a number of  oscillations with similar amplitudes can only be observed. The difference 
between focusing spectra of the armchair on one hand and of the zigzag (and reczag) edges 
on the other hand is  even more noticeable  in stronger magnetic fields, 
i.e. for $B/B_{focus}>7$. (Eventually, at very strong magnetic fields the cyclotron radius 
of the quasiparticles would become comparable to  the width of the collector and the system 
would therefore be equivalent to a Hall bar. We will not consider this regime here.)    
Note, that for a typical electron density of  $n_e=2.5\times
10^{16}\frac{1}{m^2}$ and assuming $L=1\mu m$ 
the focusing field is  $B_{focus}\approx 0.37T$, which is experimentally
feasible. 

In the case of the calculations presented in Fig.~\ref{fig:focusing-main}  the width $W$ of 
the nanoribbon was larger than $L/2$ ($L$ is the injector-collector distance). This meant that
for the Fermi energy we used  already the $p=1$ focusing peak could be observed. 
For narrow ribbons however it  can easily happen that for  given $E_F$ and $W$ 
the first few peaks would not appear, until for sufficiently strong field
the cyclotron radius $R_c$ becomes smaller than $W$ and the condition $2*p*R_c=L$ is met for some integer $p$.  
For narrow zigzag and reczag ribbons therefore there might be no observable focusing peaks
because for the magnetic fields where $R_c$ would become small enough, the $\kappa\gg 1$ condition 
[see below Eq.~(\ref{eq:zizgag-foc-cond})] is no longer satisfied. There could be of course oscillations
or peaks in the transmission  even in  this case but those would not have the simple periodicity that
focusing peaks have.

\section{Details of the numerical calculations}
\label{sec:num-calc}

Let us now briefly discuss some  of the details  of our numerical calculations for
the transmission probability $T(B)$ shown in Fig.~\ref{fig:focusing-main}. 
The injector and the collector were modeled by heavily doped graphene 
and the transmission was calculated employing the Green's function technique of
Ref.~\onlinecite{sanvito}. 
The graphene nanoribbons are assumed to be perfectly ballistic and infinitely long. This means that 
the left and right ends of the nanoribbons act as drains which absorb any particles exiting
to the left of right.  
To simulate the effect of finite temperatures we used a  simple energy averaging procedure in the 
calculation of the transmission curves:
$T(B)=\int T(B,E)\left(-\frac{\partial f_0(E)}{\partial E}\right) dE$ where 
$f_0(E)$ is the Fermi function. 
The actual results shown in Fig.~\ref{fig:focusing-main} were calculated at $T=1K$ temperature.
As it can be expected, higher temperatures tend to smear the curves while at lower ones an
additional fine structure appears.

For simple armchair and zigzag nanoribbons we assumed that the 
hopping parameter between the atoms is the same everywhere on the ribbons. In contrast, 
for the zz(57) edge we took into account that the hopping parameter changes on the pentagons
and heptagons at the edges with respect to its the bulk value.
\begin{figure}[htb]
\includegraphics[scale=0.25]{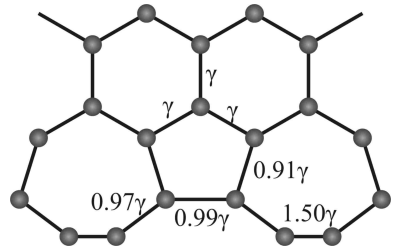}
\caption{The reconstructed zigzag  edge. The numbers indicate the change of the 
hopping parameter on particular bonds with respect to its bulk value,  $\gamma=-3.39eV$.  
\label{fig:reczag-params}} 
%\vspace{-5mm}
\end{figure}
We calculated the hoppings with density-functional tight-binding (DFTB) method, using 
the hotbit code.\cite{hotbit_wiki} This approach yields the hopping matrix elements as 
two-center integrals for pseudo-atomic orbitals that result from a straightforward pathway 
of \emph{ab initio} calculations, 
as described in Ref.~\onlinecite{koskinen_CMS_09}. The density-functional parametrization yields a valid 
description of the covalent bonding in
carbon nanomaterials.\cite{frauenheim_JPCM_02} %Fig.\ref{fig:reczag-params-and-bands}(a) 
Fig.\ref{fig:reczag-params}
shows the relative hoppings from DFTB-optimized geometry, note how the appearance of a 
triple bond in the armrest 
parts affects also the $\pi$-electron hopping, due to the significant reduction in bond length.\cite{koskinen}
\begin{figure}[hbt]
\includegraphics[scale=0.2]{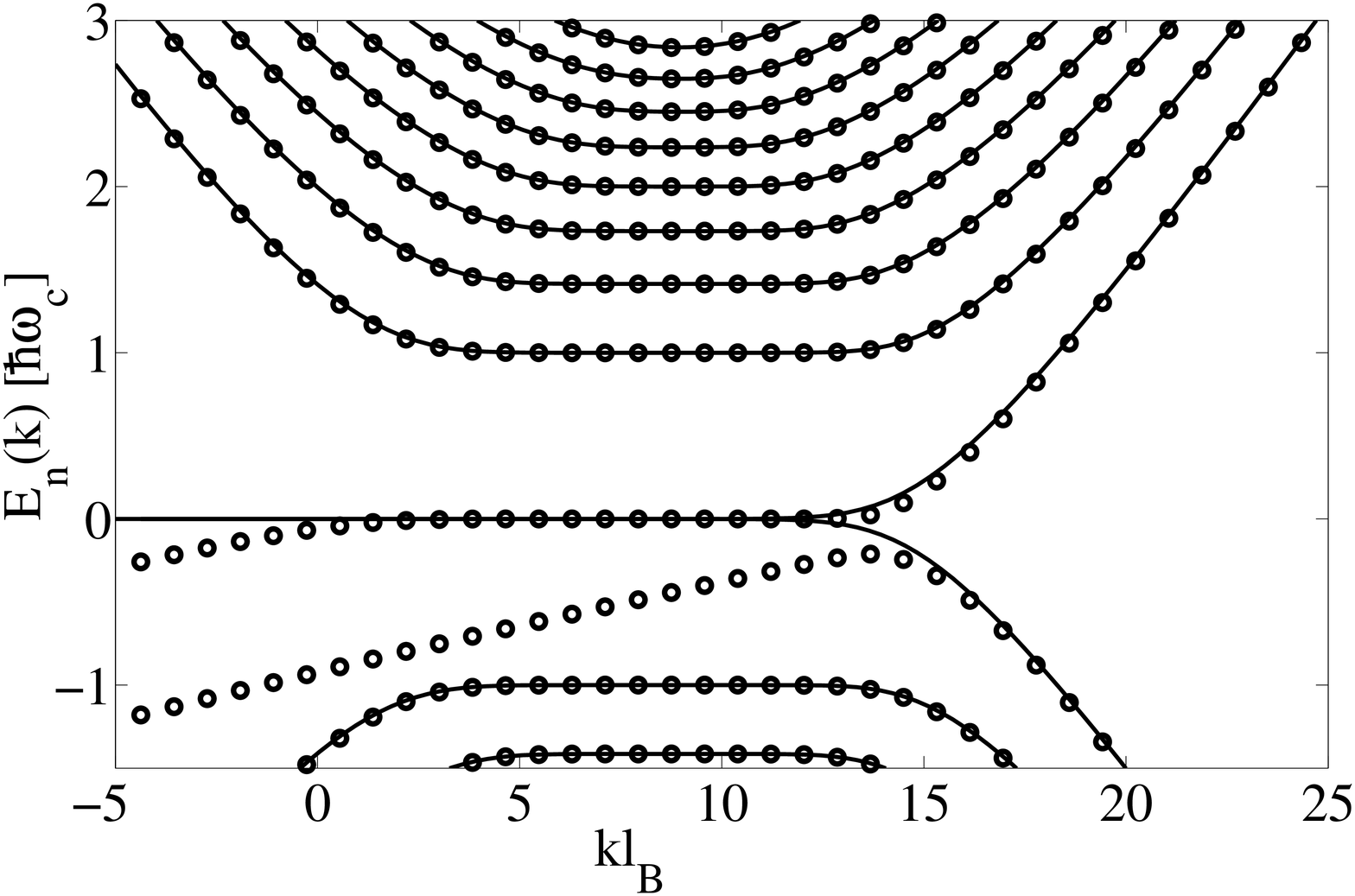}
\caption{Comparison of the band structures of zigzag (solid line) and reczag (circles)
nanoribbons from tight binding calculations 
at the K point. The strength of the  magnetic field is given by $W/l_B=9.14$.
\label{fig:reczag-bands}} 
%\vspace{-5mm}
\end{figure}
In the one orbital per site approximation 
we employed here the band structures of zigzag and reczag 
nanoribbons, apart from the close vicinity of the Dirac point,  
are very similar (see Fig.~\ref{fig:reczag-bands}). This explains why
they have very similar focusing spectra as well.

\section{Summary}
\label{sec:summary}

In summary, we  studied  coherent electron
focusing in graphene nanoribbons using both exact quantum calculations 
and in semiclassical approximation. We  found that in the case of  
armchair edges the transmission peaks are at integer multiples of  
$B/B_{focus}$ whereas for  zigzag edges such a simple rule holds only
when the filling factor $\kappa$ is much larger than unity. For zigzag nanoribbons in 
stronger magnetic field, when the filling factor is of the order of one, 
a more complex interference pattern can be observed in the transmission and  the emergence of 
this interference pattern can be understood from semiclassical calculations. 
The presence of focusing peaks at low magnetic fields can therefore attest to the 
quality of the edge structure,  while  measurements at stronger  fields  can 
discriminate between armchair and zigzag edges.
Our numerical calculations on zz(57) edges suggest that the above conclusion holds 
even if the zigzag edge is structurally reconstructed.  
Although we considered in our analytical calculations nanoribbons whose both edges had
perfect armchair or zigzag structure, we expect that our findings are more 
generally valid. Namely, they rely 
on the properties of edges states localized at one of the edges therefore
imperfections of the other edge should not affect the focusing peaks. Therefore 
this technique should be applicable to study the edges of graphene flakes as well.
Finally, an interesting extension of our work would be to consider the focusing
spectrum of other proposed\cite{wassmann} edge types 
and  possibly taking into account more than one orbital per site in the transport
computations.


\begin{thebibliography}{10}
\bibitem{parameters}
For the calculations shown in Fig.\ref{fig:focusing-main} we considered
graphene 
nanoribbons for which $W=221nm$ and 
$L=436nm$. The width of the injector and collector probes were 
$W_{I,C}=10nm$ and we assumed that the electron density in the ribbon was
$n_e=5.98\times 10^{16}\frac{1}{m^2}$. 
Although the limitations of our numerics did not allow us to consider
experimentally more realistic 
cases (i.e. $L\approx 1-2\mu m$ and $W_{I,C}\approx 50-90 nm$) we are confident
that our main results are robust and can therefore have experimental relevance.
\end{thebibliography}
\end{document}